\documentclass{emulateapj}

\usepackage{amsmath}
\usepackage{natbib}
\usepackage[colorlinks=true,citecolor=blue,urlcolor=magenta,breaklinks]{hyperref}
\usepackage{cool}
\usepackage{eulervm}
\usepackage{times}
\begin{document}
\title[The induced surface tension contribution for the equation of state of neutron stars]{
The induced surface tension contribution\\
for the equation of state of neutron stars}	

\author{Violetta V. Sagun~\altaffilmark{1,2}, Il\'\i dio Lopes~\altaffilmark{1,3}, Aleksei I. Ivanytskyi~\altaffilmark{4,2}}	
\email{violettasagun@tecnico.ulisboa.pt}
\email{ilidio.lopes@tecnico.ulisboa.pt}
\email{oivanytskyi@usal.es}
\altaffiltext{1}{
Centro de Astrof\'{\i}sica e Gravita\c c\~ao  - CENTRA,
Departamento de F\'{\i}sica, Instituto Superior T\'ecnico - IST, Universidade de Lisboa - UL, Av. Rovisco Pais 1, 1049-001 Lisboa, Portugal} 
\altaffiltext{2}{Bogolyubov Institute for Theoretical Physics, Metrologichna str. 14$^B$, Kyiv 03680, Ukraine} 
\altaffiltext{3}{Institut d'Astrophysique de Paris, Sorbonne Universit\'e,  98 bis Boulevard Arago, Paris F-75014, France}
\altaffiltext{4}{Department of Fundamental Physics, University of Salamanca, Plaza de la Merced s/n E-37008, Spain}

\date{\today}

\begin{abstract} 

We apply a novel equation of state (EoS) that includes the surface tension contribution induced by interparticle interaction and asymmetry between neutrons and protons, to the study of neutron star (NS) properties. This elaborated EoS is obtained from the virial expansion applied to multicomponent particle mixtures with hard core repulsion. The considered model is in full concordance with all the known properties of normal nuclear matter, provides a high-quality description of the proton flow constraints, hadron multiplicities created during the nuclear-nuclear collision experiments, and equally is consistent with astrophysical data coming from NS observations. The analysis suggests that the best model parameterization gives the incompressibility factor $K_{0}$, symmetry energy $J$, and symmetry energy slope $L$ at normal nuclear density equal to $200$ MeV, $30$ MeV, and $113.28-114.91$ MeV, respectively. The mass-radius relations found for NSs computed with this EoS are consistent with astrophysical observations.

\end{abstract}


\section{Introduction}
Compact stars are the most exotic and challenging laboratories in the universe to test the properties of baryonic matter. Inside these objects, ultra-cold matter is known to be in the most extreme conditions characterized by very high baryonic densities, rotation speeds, and strong magnetic fields \citep{2010Natur.467.1081D, 2013Sci...340..448A, 2016ARA&A..54..401O}. Understanding the complex physical phenomena occurring inside such compact astrophysical objects as neutron stars (NSs), hybrid stars (HSs), and quark stars (QSs) requires a profound knowledge of a wide range of scientific disciplines in physics, as well as an analysis of a large amount of observational and experimental data. The main goal of the present research is to obtain an accurate equation of state (EoS) able to describe strongly interacting matter through a wide range of baryonic densities. Although the specific properties of the matter inside compact objects remain poorly known, nucleus-nucleus (A+A) collision experiments have provided a way to study the properties of strongly interacting matter in extreme conditions. These experimental programs are so far the best terrestrial laboratories to study nuclear matter at high density.

\paragraph*{}
On the other hand, the astronomical study of strongly interacting matter has received a fresh kick with the recent observation of a gravitational-wave (GW) signal coming from the first NS-NS binary merger ever detected -- the event GW170817~\citep[i.e.,][]{2017PhRvL.119p1101A}. Moreover, this GW event was also followed by a large set of astronomical observations across the electromagnetic spectrum~\citep[i.e.,][]{2017ApJ...848L..12A}. This combined set of observations has opened a unique new window into the study of the properties of matter inside NSs. \citet{2011PhRvD...83..124008H}, among other authors, has shown that the specific dynamic properties of the NS-NS binary merger depend strongly on the softness or stiffness of the EoS. Indeed, the GW data, as well as the multi-messenger observations of the NS-NS binary coalescence, are very sensitive to the specific properties of the EoS~\citep{2018RPPh...81e6902B}.  Moreover, we expect that in the near future it will be possible to put even more stringent constraints on the EoS, with the discovery of new NS-NS binary mergers and even more so with the yet undiscovered NS-black hole binary mergers~\citep{2011LRR....14....6S}.   

\paragraph*{}
Still, modern A+A collision programs provide us with the only way to obtain detailed information about the properties of nuclear and hadron matter in a controlled experimental environment. This class of studies complements and challenges  the multi-messenger observations coming from compact star binary mergers such as NS-NS, NS-black hole, or other more exotic binary compact object mergers with HSs or QSs as a component{s}. In A+A collisions matter, can be compressed up to particle densities achieved inside NSs, but with temperatures much higher than the ones expected inside these objects. 

Indeed, the final stage of the least energetic A+A collisions is characterized by thermal excitation energies per particle well above $50~MeV$,  which is equivalent to $2\cdot10^8~K$ and significantly exceeds temperatures of equilibrated NSs of about $10^6~K$ \citep{2017PhRvD..96d3015C}. Even huge temperatures of $10^{11}~K$ typical for proto-NS are small compared to $6\cdot 10^{12}~K$ achieved recently in A+A collisions at ALICE CERN.

These essential differences significantly complicate the study of the baryonic matter EoS and disable its direct extrapolation from A+A collision regions of strongly interacting matter phase diagram to regimes typical for NSs. As previously mentioned, experimental data on A+A collisions are so far the best available source of information about the properties of nuclear matter in extreme conditions. Thus, analysis of these data is of high interest for modeling the EoS of dense hadronic matter, which is an important input for astrophysical applications, such as the ones mentioned above.

\paragraph*{}
Currently, astrophysical observations have been very useful to put strong constraints on the EoS of strongly interacting matter at zero temperature \citep{2010Natur.467.1081D, 2013Sci...340..448A}. Different data sets have inferred slightly different constraints on the values of NS radii.  ~\citet{2016ARA&A..54..401O} put constraints on the mass-radius relation using data on bursting NSs that show photospheric radius expansion, while ~\citet{2010ApJ...722...33S, 2013ApJ...765L...5S} did the same analysis both for transiently accreting and bursting sources. Using more than $10^9$ equilibrium models and tidal deformability from the GW170817 signal ~\citep{2017PhRvL.119p1101A}, another radius constraint for a $1.4M_{\odot}$ star was found by ~\citet{2018arXiv180300549M}. These constraints can be satisfied only if, at densities below two saturation densities, the EoS is relatively soft, while higher densities lead to its fast stiffening. The latter feature of an EoS is required in order to provide its consistency with the highest observed NS mass equal to 2.01(4) solar masses \citep{2013Sci...340..448A}.  Such a behavior of an EoS corresponds to the sudden increase of the sound speed. This can even lead to its superluminal values, and consequently, to a violation of the causality, which, however, has to be respected by a realistic EoS. Together with the constraints coming from heavy-ion collision experiments, which determine the properties of the strongly interacting matter at high temperatures and low baryonic densities, this imposes strong limitations on an EoS.

\paragraph*{}
The formulation of a realistic EoS is a highly non-trivial task, which requires a delicate balance between the repulsion and attraction of the constituent particles \citep{2012ARNPS..62..485L} and simultaneous fulfillment of the mentioned astrophysical and A+A collision constraints. In addition, a realistic EoS has to reproduce the saturation properties of the symmetric nuclear matter in the ground state. The latter condition means that at zero temperature and normal nuclear density $n_{0} \simeq 0.16~fm^{-3}$ symmetric nuclear matter has zero pressure and binding energy per nucleon that is approximately equal to $16~MeV$ ~\citep{1971ARNPS..21...93B}. Another crucial parameter that represents the softness/stiffness of the EoS is the incompressibility factor $K_{0}=9\frac{\partial p~}{\partial n_B}$ at saturation, where $p$ and $n_B$ are the system pressure and baryonic density, respectively. Recent analyses of the experimental data indicate the value  $K_0=230 \pm 30$  MeV ~\citep{2009PhRvC..80a1307K, 2012PhRvC..85c5201D}. Moreover, these values of $K_0$ are in full agreement with a proton flow constraint \citep{2002Sci...298.1592D}.

NS properties are found to be strongly sensitive to the symmetry energy $E_{sym}$, which is equal to the difference between energy per particle in symmetric and pure neutron matter ~\citep{2006PhRvC..74c5802K, 2014PhRvC..90d5802B}. Its slope $L=3n\frac{\partial E_{sym}}{\partial n}$ at saturation is an important indicator of the symmetry energy stiffness  at high densities. These two characteristics of the nuclear EoS influence the NS radius, the composition of the crust, and a transition between crust and core. Nuclear experiments suggest the value of symmetry energy at saturation $E_{sym}(n=n_{0})\equiv J=30\pm 4$ MeV, while the extracted $L$ varies in the range between 20 and 115 MeV, depending on chosen observables and analysis methods ~\citep{2013PhLB..726..234Z}. 

At present, quantum chromodynamics (QCD) cannot provide us with information about a strongly interacting matter EoS at high baryonic densities. This requires the use of the phenomenological models. At the same time, models of this type have to be consistent with the conceptual fact of deconfinement of quarks and gluons at sufficiently high densities and/or temperatures. Hard core repulsion is one of the most successful approaches to suppress hadronic excitations and provide  formation of the deconfined phase of QCD ~\citep{1991ZPhyC..51..485,2012LNP...841.....S}. Moreover, hard core repulsion of hadrons is necessary to reproduce the multiplicities of these particles measured in A+A collisions ~\citep{2006NuPhA.772..167A}. As was recently shown by ~\citet{2017JPhCS.779a2012A} and ~\citet{2017arXiv170300049S} experimental data prefer the hard core radius of baryons in the range between $0.3~fm$ and $0.5~fm$, respectively. Such a difference in the baryon radius values is caused by the simple fact that ~\citet{2017JPhCS.779a2012A} used the model where all particles are characterized by the same radius. This approach prefers intermediate values between the small radii of $\pi$-mesons (being the most abundant particles) and the large radii of baryons.

\paragraph*{}
The previous formulations of the hadronic matter EoSs with the hard core repulsion between particles (\citet{2006NuPhA.772..167A, 2012arXiv1204.0103O}) have a relatively small applicability range because they are based on the Van der Waals approximation. These difficulties were overcome by a recently proposed EoS that includes the induced surface tension (IST) generated by the interparticle interaction, the so-called IST EoS~\citep{2014NuPhA.924...24S}. Contrary to the eigensurface tension caused by the pulling constituents at the surface inward due to their attractive interaction, the IST is generated by the repulsion between outer and surface constituents, which also pushes the latter ones inward.
The medium effects accounted via the IST allow one to go beyond the usual Van der Waals approximation ~\citep{2017arXiv170300049S}, which increases the EoS range of causality and applicability. As shown by \citet{2017EPJWC.13709007S} this theoretical framework makes it possible to describe the A+A collision data measured at Alternating Gradient Synchrotron (AGS), Super Proton Synchrotron (SPS), Relativistic Heavy Ion Collider (RHIC), and Large Hadron Collider (LHC) energies ~\citep{2017arXiv170300049S}, nuclear matter properties ~\citep{2014NuPhA.924...24S}, and compact astrophysical objects~\citep{2017arXiv170907898S}. 

Moreover, the realistic EoS that describes strongly interacting matter properties at low temperature should be able to correctly reproduce the nuclear liquid-gas phase transition with physically correct properties of its critical endpoint \citep{2017arXiv171008218I}. To describe the full set of constraints, a new family of generalized IST EoSs was proposed by ~\citet{2017arXiv171008218I}. This approach also treats the attraction between the particles in a more realistic way compared to ~\citet{2017arXiv170907898S}.

\paragraph*{}
The goal of the present work is to extend the novel generalized IST EoS \citep{2017arXiv171008218I} to account for an asymmetry between neutrons and protons to the description of matter inside the NSs. We tune the present model parameters to make it consistent with the A+A collision data on yields of baryons and to simultaneously describe with high precision the experimental flow constraint, as well as to reproduce the mass-radius relation of NSs obtained from astronomical observations. This allows us to fix an EoS that successfully describes nuclear and hadronic matter in all accessible intervals of densities and temperatures. We show the exceptional role of the IST in accurate description of strongly interacting matter inside NSs.
 
\paragraph*{}
The proposed IST EoS is the first step toward the development of a unified model able to describe strongly interacting matter in all ranges of temperature and baryon density. To describe core-collapse supernovae, and proto-NSs, as well as experimental data on A+A collisions, the IST approach developed for zero temperature in the present work can be generalized to finite temperature. An additional advantage of the IST EoS is that it can be extended for an arbitrarily large number of different particle species, as it was done in \citet{2018NuPhA.970..133B}.

\paragraph*{}
This article is organized as follows. In Sec. \ref{sec-2} we briefly review the generalized IST EoS and show the effects of the IST term. Sect. \ref{sec-3} is devoted to the determination of the model parameters and to calculation of the NS mass-radius relation. The paper ends with a summary in Sect. \ref{sec-4}.

\section{EoS: model description}
\label{sec-2}
In this work we use the multicomponent version of the quantum generalization of the IST EoS with the mean field interaction between particles  (see ~\citet{2014NuPhA.924...24S} and ~\citet{2017arXiv171008218I} for details). The NS matter is considered as a mixture of neutrons, protons, and electrons (subscript indexes $``n"$, $``p"$, and $``e"$, respectively) with corresponding masses $m_A$ and chemical potentials $\mu_A$ ($A=n,~p,~e$). Non-strange heavy baryons like $\Delta$-isobars are not included in the present EoS due to their small fractions reported by ~\citet{2017NuPhA.961..106K}. At the same time, the contribution of mesons is suppressed at zero temperature because they are bosons. We neglect the Coulomb interaction of electrically charged particles and treat electrons as non-interacting and point-like objects with a zero hard core radius $r_e=0$. This approximation is valid for the conditions found in cold NSs, where EoS properties are mostly defined by the strong interactions between protons and neutrons. These particles are supposed to have the same hard core radii $r_n=r_p=r$, which is justified by the fit of A+A collision experimental data with the multicomponent IST EoS ~\citep{2018NuPhA.970..133B}. In the present work we treat $r$ as a free parameter in the range from 0.3 to 0.5 fm.  We neglect the electron mass ($m_e=0$), while neutrons and protons are supposed to have the same masses $m_n=m_p=m$. The Grand Canonical Ensemble phenomenological EoS of the neutron-proton-electron mixture has the form of the system of two coupled equations for the pressure $p$ and the IST coefficient $\sigma$. Thus,
\begin{eqnarray}
\label{Eq1}
p&=&
\hspace*{-.3cm}\sum_{A=n,p,e}p^{id}(m_A, \nu^1_A)
-p^{int}(n^{id}_B)+p^{sym}(n^{id}_B,I^{id}) \,,\quad\\
\label{Eq2}
\sigma&=&\hspace*{-.2cm}\sum_{A=n,p}p^{id}(m_A, \nu_A^2) r \, .
\end{eqnarray}
Partial pressures and the IST coefficients of neutrons, electrons, and protons in  Eqs. (\ref{Eq1}) and (\ref{Eq2}) are written in terms of the zero temperature pressure $p^{id}$ of non-interacting Fermi particles with spin $\frac{1}{2}$ and quantum degeneracy $2$, thus 
\begin{eqnarray}
\label{Eq3}
p^{id}(m,\mu)&=&
\frac{\mu k(2\mu^2-5m^2)+3m^4\ln\frac{\mu+k}{m}}{24\pi^2}\theta(\mu-m),\quad
\end{eqnarray}
where $k=\sqrt{\mu^2-m^2}$ is the Fermi momentum of a particle with mass $m$ and chemical potential $\mu$, and  $\theta$ is the Heaviside function. It is worthwhile to note that electrons do not contribute to the IST coefficient because $r_e=0$. The mean field interaction between neutrons and protons is accounted for by the term $p^{int}$, which is defined through the baryon charge density of ideal gas $n^{id}_B=n^{id}(m,\nu^1_n)+n^{id}(m,\nu^1_p)$ with
\begin{eqnarray}
\label{Eq4}
n^{id}(m,\mu)&=&
\frac{\partial p^{id}(m,\mu)}{\partial\mu}=\frac{k^3}{3\pi^2}\theta(m-\mu) \,
\end{eqnarray}
being the density of non-interacting fermions with spin $\frac{1}{2}$. In the present work we generalize the IST EoS developed by \citet{2017arXiv171008218I} to the case of finite asymmetry between nucleons and protons, which gives rise to the symmetry energy correction $p^{sym}$ in Eq. (\ref{Eq1}). The dependence of this term on the baryon charge density $n^{id}_B$ and asymmetry parameter $I^{id}=(n^{id}(m,\nu^1_n)-n^{id}(m,\nu^1_p))/n^{id}_B$ of ideal gas is discussed below.

Physical chemical potentials of  neutrons and protons $\mu_A$ are changed to the effective ones: $\nu_A^1$ and $\nu_A^2$ with $A=p,n$. These effective chemical potentials incorporate the effects of the hard core repulsion through the nucleon eigenvolume ${\rm V}=\frac{4}{3}\pi r^3$ and surface $S=4\pi r^2$, whereas the mean field attraction and symmetry energy are accounted for through the density and nucleon asymmetry dependent potentials $U$ and $U^{sym}$, respectively~\citep{1988JPhG...14..191R}. Thus,
\begin{eqnarray}
\label{Eq5}
\nu^1_A&=&\mu_A-p{\rm V}-\sigma S+U(n^{id}_B) \mp U^{sym}(n^{id}_B,I^{id})\,,\\
\label{Eq6}
\nu_A^2&=&\mu_A-p{\rm V}-\alpha\sigma S+U_0\,,
\end{eqnarray}
where the signs ``-'' and ``+''  correspond to neutrons and protons, respectively, $U_0$  is a constant parameter that accounts for the mean field  effects and $\alpha$ is a dimensionless parameter with a value always larger than one \citep[$\alpha>1$,][]{2014NuPhA.924...24S}. \citet{2017EPJWC.13709007S} has shown that in the Boltzmann statistics case, $\alpha =1.245$ reproduces the correct values of the third and the fourth  virial coefficients of the gas of hard spheres. 
Additional justification of this value of $\alpha$ was
independently found by \citet{2017arXiv170300049S}. It was shown in that paper that values very close to $\alpha=1.25$ generate the widest causality range of the Boltzman mixture of particles. Because for the quantum case, the higher virial coefficients are not known, in this work we performed an independent check of which value of $\alpha$ is preferred by astrophysical data. Our analysis demonstrates that a realistic dependence of the total NS mass on its radius is obtained for $\alpha$ from 1.2 to 1.3. At the same time, the $\alpha=1.1$ also reported in ~\citet{2017arXiv170907898S} is not consistent with a realistic mass-radius diagram. This allows us to conclude that the contribution coming from  the nuclear asymmetry energy with realistic values of $J$ and $L$ narrows a range of possible values of $\alpha$. Therefore, we set this parameter to be equal to 1.245. For unification of notations it is convenient to introduce the effective chemical potentials of electrons defined as $\nu^1_e=\mu_e$ and $\nu^2_e=0$. 

The thermodynamic consistency of the model with the mean field interaction requires a special relation between the interaction pressure $p^{int}$ ~\citep{1988JPhG...14..191R, 1989ZPhyC..43..261B, 2017arXiv170406846B, 2017arXiv171008218I} and potential $U$, that reads as
\begin{eqnarray}
\label{Eq7}
p^{int}(n^{id}_B)=\int\limits_0^{n^{id}_B}dn~n~\frac{\partial U(n)}{\partial n}\,.
\end{eqnarray}
Following  \citet{2017arXiv171008218I}, the parameterization of the interaction potential was chosen to be  $U(n)=C_d^2 n ^{\kappa}$, which gives us the opportunity to calculate the interaction term for pressure as $p^{int}(n)= \frac{\kappa}{\kappa +1}C_d^2 n ^{\kappa +1}$. The consistency condition (\ref{Eq7}) also explains the absence of the mean field contribution in the expression for the IST coefficient in Eq. (\ref{Eq2}). Indeed, according to Eq. (\ref{Eq6}), the mean field contribution to the effective chemical potentials of nucleons obtained from the $\sigma$ expression is $U_{0}=const$. Substituting this constant potential into Eq. (\ref{Eq7}) we immediately obtain zero. The requirement of thermodynamic consistency also relates the symmetry energy pressure and the corresponding interaction potential. At the same time, $p^{sym}$ and $U^{sym}$ depend on two quantities, namely on $n^{id}_B$ and $I^{id}$. Therefore, in this case the thermodynamic consistency requires simultaneous fulfillment of two relations similar to Eq. (\ref{Eq7}). These relations are
\begin{eqnarray}
\label{Eq8}
p^{sym}(n^{id}_B,I^{id})&=&I^{id}\int\limits_0^{n^{id}_B}dn~n~\frac{\partial U^{sym}(n,I^{id})}{\partial n}\, ,
\end{eqnarray}
and
\begin{eqnarray}
\label{Eq9}
p^{sym}(n^{id}_B,I^{id})&=&n^{id}_B\int\limits_0^{I^{id}}dI~I~\frac{\partial U^{sym}(n^{id}_B,I)}{\partial I}\,.
\end{eqnarray}
Conditions (\ref{Eq8}) and (\ref{Eq9}) are satisfied simultaneously only if arguments of the symmetry energy pressure and corresponding interaction potential enter on these relations as a product. In other words, 
$p^{sym}=p^{sym}(n^{id}_BI^{id})$ and $U^{sym}=U^{sym}(n^{id}_BI^{id})$, thus
\begin{eqnarray}
\label{Eq10}
p^{sym}(n^{id}_BI^{id})&=&\int\limits_0^{n^{id}_BI^{id}}dn~n~\frac{\partial U^{sym}(n)}{\partial n}\,.
\end{eqnarray}
In this work we parameterize the symmetry energy pressure as $p^{sym}(n)=\frac{A^{sym}n^2}{1+(B^{sym}n)^2}$,
where $A^{sym}$ and $B^{sym}$ are constants fitted to the values of $J$ and $L$. This also makes it possible to unambiguously define $U^{sym}$. Note that such a parameterization of the asymmetry energy is not unique. It provides a correct quadratic dependence of $p^{sym}\sim {I^{id}}^2$ on the asymmetry parameter at low baryonic densities, while at high ones symmetry energy pressure approaches the limiting value $\frac{A^{sym}}{(B^{sym})^2}$.

\begin{table*}[t]
\caption{\label{tab1} IST EoS  parameters}
\begin{center}
\vspace*{-0.5cm}
\begin{tabular}{|c|l|l|l|l|l|l|l|l|l|l|l|}
\hline	
IST EoS\footnote{\footnotesize
Parameters of the IST EoS, which simultaneously reproduce the nuclear matter ground-state properties and the flow constraint, and generate a realistic critical endpoint of nuclear matter. Adjustable parameters of the model include $r$, $\alpha$, $\kappa$, $B^{sym}$, $A^{sym}$, $C_d^2$, and $U_0$, whereas $K_0$, $J$, $L$, and $M_{max}$ are calculated for each of their sets.}    
& $r$ & $\alpha$ & $\kappa$ & $B^{sym}$& $A^{sym}$ &$C_{d}^{2} $ & $U_0$  & $K_0$  &  $J$ & $L$  & $M_{\footnotesize max}$\\
{\footnotesize Sets} &  $(fm)$  & $-$ & $-$  & $(fm^{3})$ & $\footnotesize (MeV \cdot fm^{3})$& $\footnotesize (MeV \cdot fm^{3\kappa})$& (MeV) & (MeV) & (MeV)  & (MeV) & $(M_{\odot})$ \\
\hline
A (magenta curve on Fig. \ref{fig:b})& 0.492  & 1.245  & 0.263 & 3.5 & 16.896 & 143.564 & 147.456 &  200.03 & 30.0 & 114.91 & 2.229 \\
\hline
B (blue curve on Fig. \ref{fig:b})& 0.484  & 1.245  & 0.26 & 4.5 & 14.762 & 144.042 & 150.97 &  200.00 & 30.0 & 113.28 & 2.189 \\
\hline
\end{tabular}
\end{center}
\end{table*}

The IST EoS given by Eqs. (\ref{Eq1}) and (\ref{Eq2}) enables us to find the particle number densities of neutrons, protons, and electrons as total derivatives of the pressure $p$ with respect to corresponding chemical potentials, i.e. $n_A=\frac{\partial p~}{\partial \mu_A}$. 
The conditions of thermodynamic consistency given by Eqs. (\ref{Eq7}) and (\ref{Eq10}) significantly simplify the calculation of these quantities, which yields
\begin{eqnarray}
\label{Eq11}
n_A&=&\frac{n^{id}(m,\nu^1_A)-
3n^{id}_BV(1-\alpha\Phi)n^{id}(m,\nu^2_A)}
{1+n^{id}_BV(1-\Phi)}\,,\end{eqnarray}
where $A=n,p,e$ and $\Phi$ is given by  
\begin{eqnarray}
\label{Eq12}
\Phi=\frac{3\sum\limits_{A=p,n}n^{id}(m,\nu^2_A)V}
{1+3\alpha\sum\limits_{A=p,n}n^{id}(m,\nu^2_A)V}\, .
\end{eqnarray}
Explicit expressions for the particle densities of neutrons and protons allow us to find the baryonic density as $n_B=n_n+n_p$. Because the NS is assumed to be electrically neutral, the total density of electric charge equals to zero, i.e. $n_Q=n_p-n_e=0$.  Furthermore, inside the NS rates of direct and inverse $\beta$-decay processes, $n\leftrightarrow p+e$ are equal, which leads to chemical equilibrium between neutrons, protons, and electrons. This equilibrium is accounted for by the condition $\mu_n=\mu_p+\mu_e$. In this study the  neutrino contribution is neglected.

The zero temperature energy density of electrically neutral equilibrated mixture of neutrons, protons, and electrons is defined as 
\begin{eqnarray}
\epsilon=\sum_{A=n,p,e}\mu_An_A-p=\mu n_B-p,
\end{eqnarray}
where $\mu=\mu_{n}$ is the baryonic chemical potential. This expression implicitly defines the dependence of the pressure on the energy density that is required to solve the Tolman-Oppenheimer-Volkoff~\citep[TOV,][]{1934rtc..book.....T, 1939PhRv...55..364T, 1939PhRv...55..374O} equation in the closed form.

In order to produce a more realistic internal structure of an NS, we introduce a polytropic EoS  that mimics the atomic structure of the outer and inner crusts on the top of the NS's core. In this stellar model the outer crust is  formed by a lattice of nuclei mix with the gas of ultra-relativistic electrons. As usual, we assume that a polytropic EoS of the form $p=K_{pol}n_{B}^{\gamma}$ is sufficient to describe its thermodynamics. Following ~\citet{2013A&A...560A..48P}, the adiabatic index $\gamma$ is chosen to be equal to $4/3$. The polytropic EoS was smoothly matched to the IST one. The  transition densities about $0.09~fm^{-3}$ and $K_{pol}$ were defined by requiring the same pressure and energy density of these EoSs at the transition between them. It is necessary to stress that due to its small value the transition density is almost the same for two sets of the IST EoS parameters from Table \ref{tab1} because at such a regime both parameterizations are weakly sensitive to the details of the interparticle interaction.
The introduction of this polytropic EoS affects the mass-radius relation only for NS masses below $0.2 M_{\odot}$ and has a negligible effect at higher masses.

\begin{figure}[!]
\centering
\includegraphics[scale=0.5]{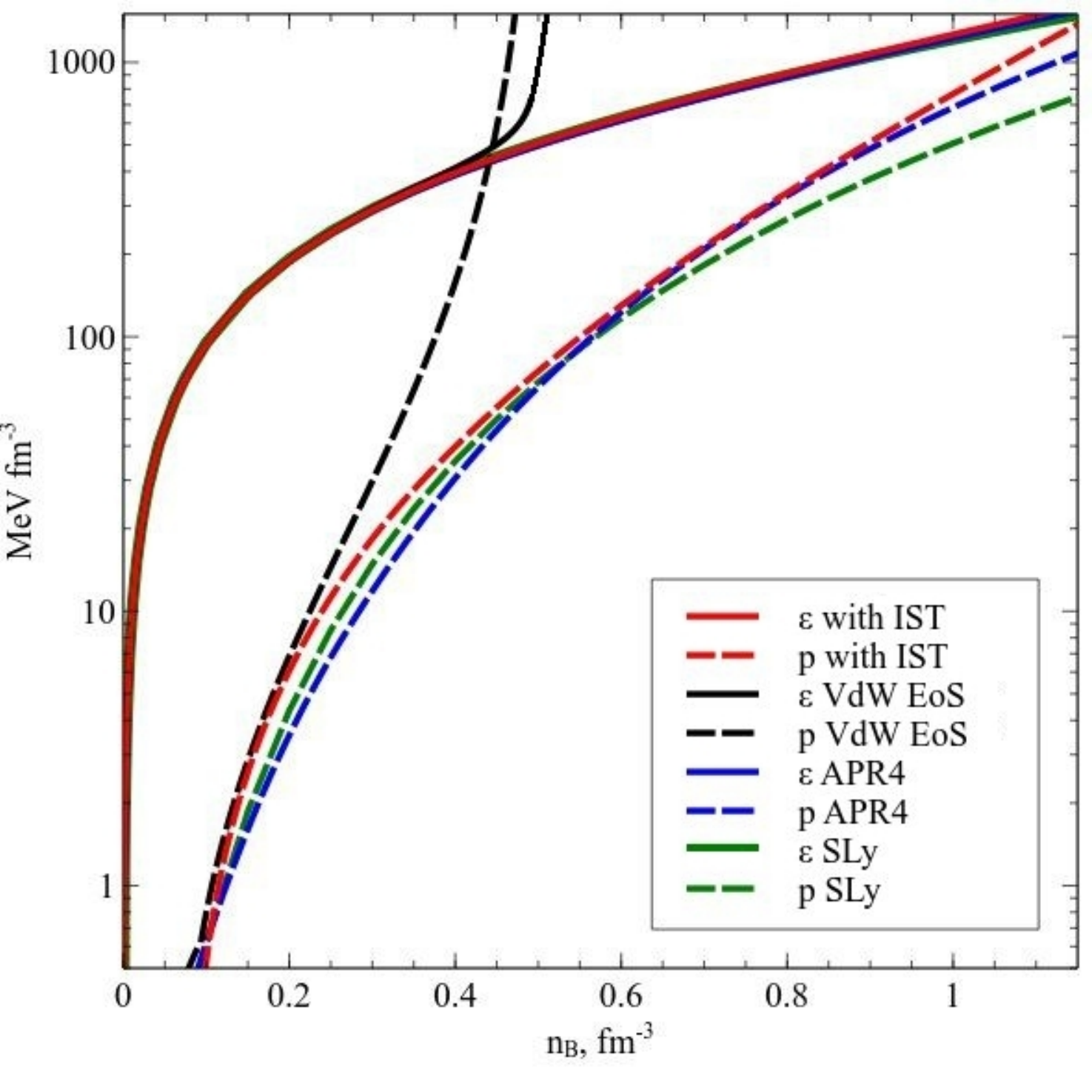}
\includegraphics[scale=0.6]{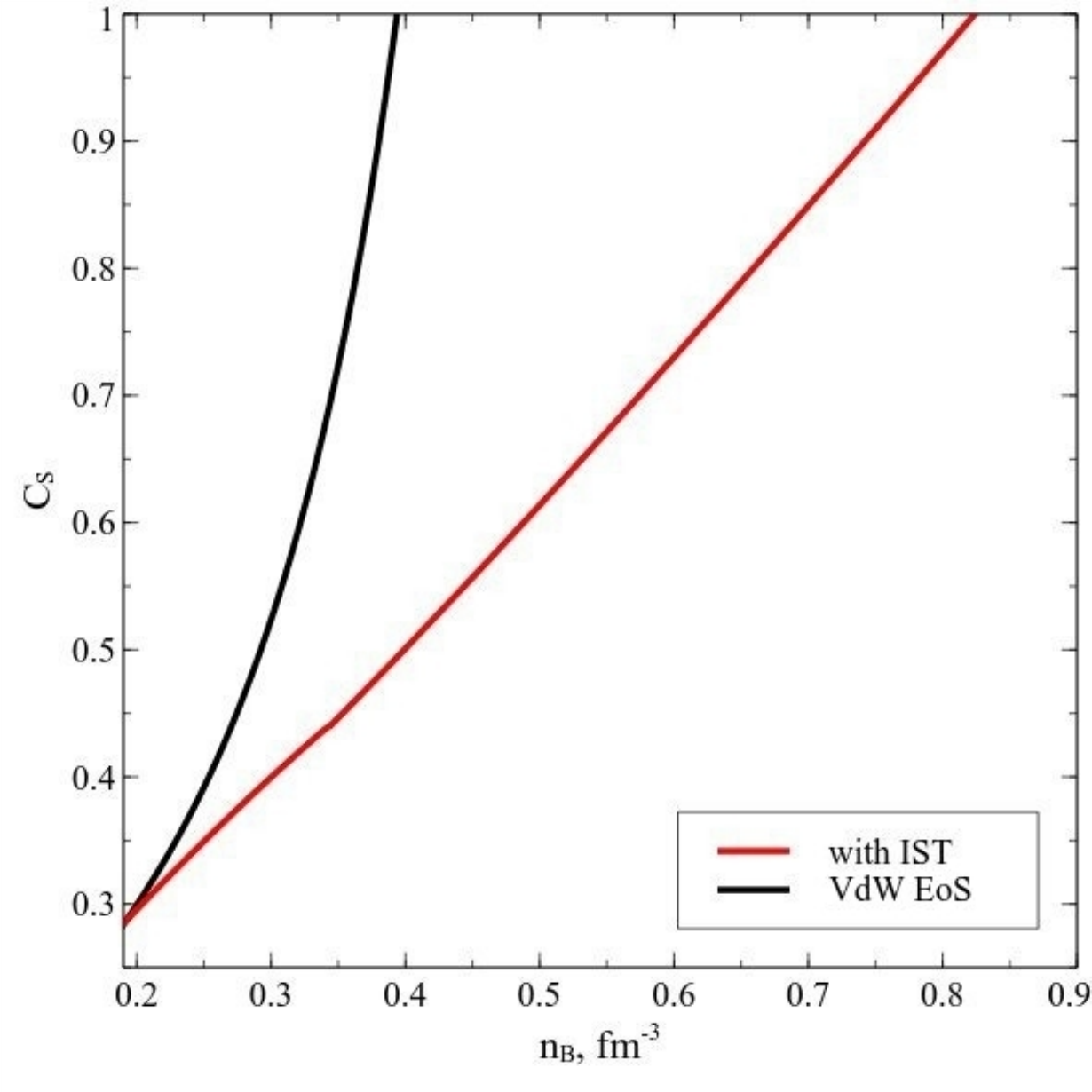}    
\caption{{Upper panel:} the energy density $\epsilon$ (solid curves) and pressure $p$ (dashed curves) as functions of the baryonic density $n_B$ in comparison to the APR4 \citep{1998PhRvC..58.1804A} and SLy \citep{2001A&A...380..151D} EoSs. {Lower panel:} speed of sound $C_s$ as functions of the baryonic density $n_B$. Calculations are performed for the set A of the model parameters (see Table \ref{tab1}) for the IST EoS (red curves) and the vdW EoS (black curves).}
\label{fig:a}
\end{figure}

\paragraph*{}
The model parameters were tuned in order to reproduce the discussed properties of nuclear matter, to be consistent with the flow constraint and experimental data on hadronic yields, as well as to simultaneously fit the astrophysical constraints on the mass and radius of NSs. Corresponding values of parameters are refereed to as set A and listed in Table~\ref{tab1}. Because the APR4, which stands for A18+$\delta$v+UIX* parameterization, \citep{1998PhRvC..58.1804A} and SLy ~\citep{2001A&A...380..151D} EoSs are usually used as references in many nuclear and astrophysical studies,  we also found an alternative parameterization of the IST EoS, which predicts an NS mass-radius relation that is close to the ones of these EoSs. This parameterization is referred to as set B in Table~\ref{tab1} and also reproduces the discussed properties of nuclear and hadronic matter. It is worthwhile to mention that in \citet{2017arXiv171008218I} it was found that the $\kappa$ values consistent with the flow constraint and realistic values of $K_0$ should lie in the range between 0.1 and 0.3. Indeed, in the present study we conclude that astrophysical data prefer $\kappa \simeq 0.26$.

As one can see from Fig. \ref{fig:a}, the IST has a huge impact on the EoS. At $n_B\simeq 0.4~fm^{-3}$ the present EoS already has essentially smaller pressure and energy density compared to the EoS obtained within the widely used Van der Waals parameterization of the hard core repulsion (see the upper panel of Fig. \ref{fig:a}). Such an EoS is referred to as the VdW one and can be formally obtained from the IST EoS by setting $\sigma=0$ and multiplying the eigenvolume $V$ by the factor 4, which ensures the correct value of its second virial coefficient. Calculation of the VdW EoS was performed for the same values of the hard core radius and parameter $\kappa$ within the framework of ~\citep{1988JPhG...14..191R,1991ZPhyC..51..485}. The low pressure of the IST EoS compared to the VdW one indicates its significant softness and wide range of causality, where the speed of sound $C_{s}^2=\frac{dp}{d \epsilon}$ does not exceed the speed of light. Indeed, as it is seen from the lower panel of Fig. \ref{fig:a}, the IST EoS remains causal up to $n_B\simeq0.83 ~fm^{-3}$ (about 5.2 normal nuclear densities), whereas the VdW EoS provides causality only up to $n_B\simeq0.4 ~fm^{-3}$ (only 2.5 normal nuclear densities). In other words, the IST effects are very important toward the NS center. Their account widens the present EoS applicability range more than twice. The comparison with the APR4 \citep{1998PhRvC..58.1804A} and SLy \citep{2001A&A...380..151D} EoSs shows almost the same behavior of pressure and energy density for the IST EoS and APR4, while the SLy EoS is a little bit softer at higher baryon densities. This allows us to conclude that the IST EoS is a powerful and universal tool not only to interpret the data coming from the collision experiments, but also to predict the properties of nuclear and hadron matter at extreme conditions and to study the compact star interiors.

\section{The mass-radius relation for an NS}
\label{sec-3} 

The total mass and radius of a given NS are obtained by integration of the TOV equation, which connects the stellar matter properties defined by the EoS and its macroscopic properties. The integration procedure is performed from the center of NS, where the energy density is fixed as an initial condition, to its surface. The latter is a sphere of radius $R$ defined by the zero pressure condition. The total NS mass $M$ is obtained by integrating $\epsilon$ over the star volume.

\begin{figure}[!]
\centering
\includegraphics[scale=0.5]{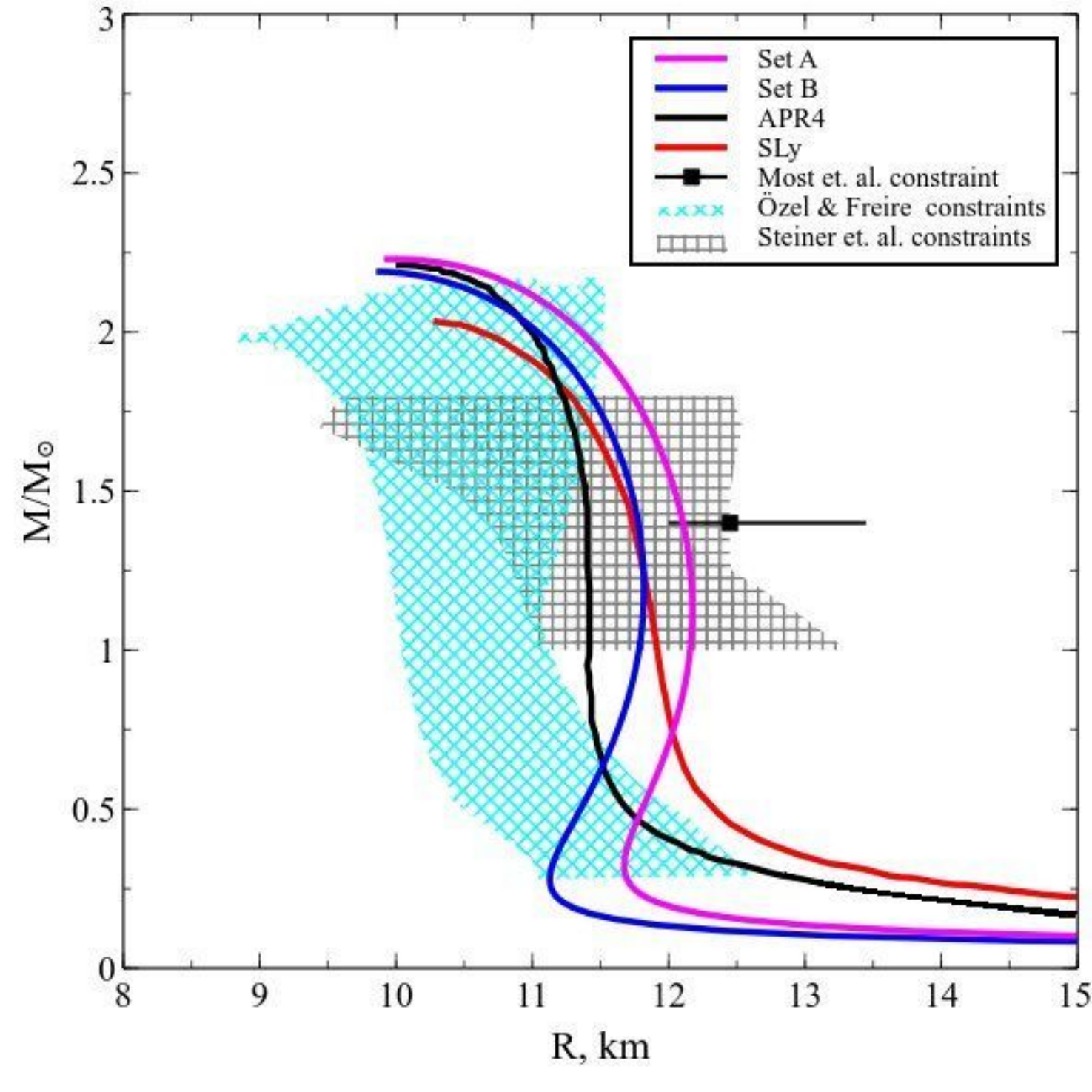}
\includegraphics[scale=0.5]{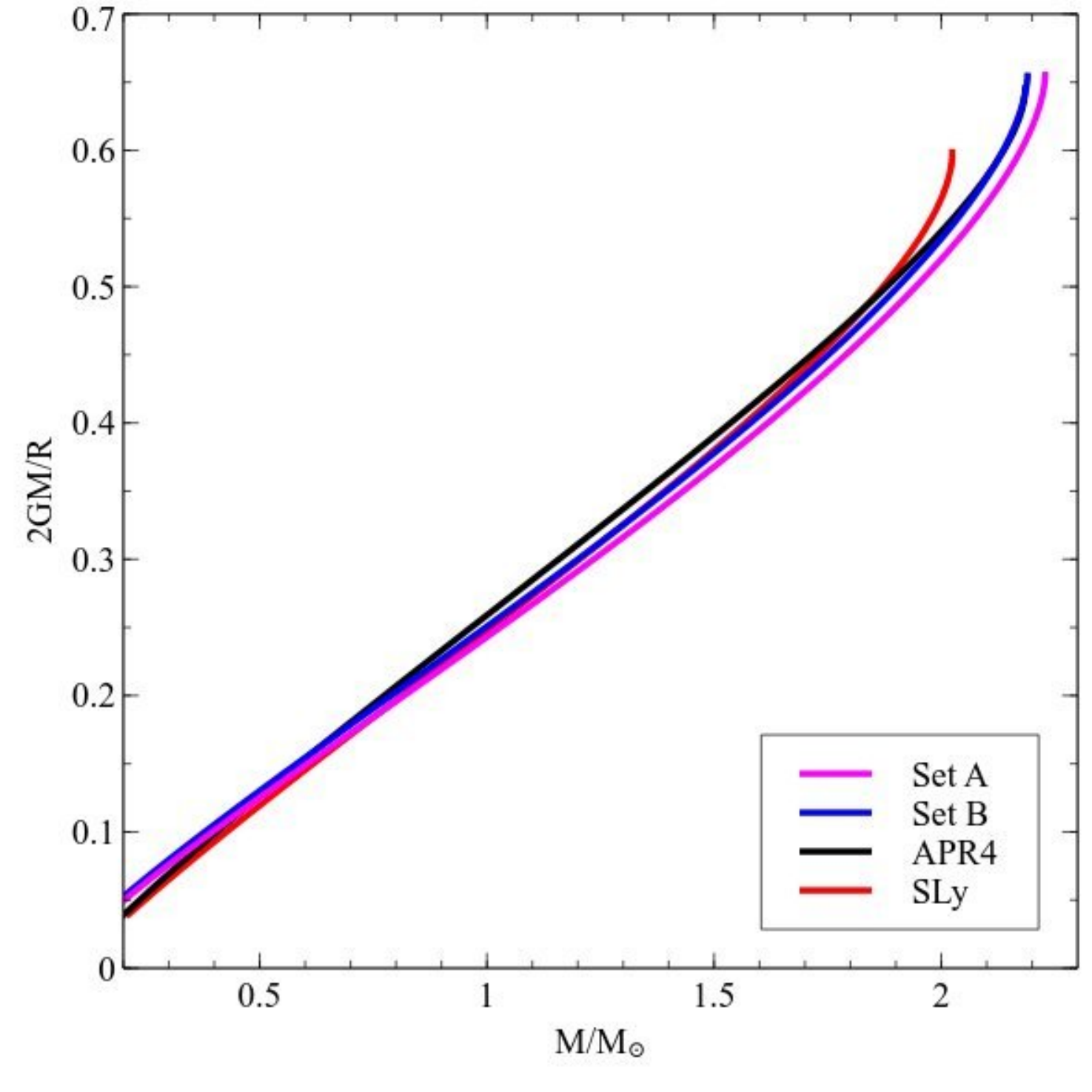}  
\caption{{Upper panel:} 
relation between gravitational mass $M$ of an NS and its radius $R$. {Lower panel:} the NS compactness $2GM/R$ as a function of its mass $M$. Calculations are performed for sets A and B of the IST EoS (see Table \ref{tab1}). Sets A and B are compared to the famous APR4  \citep{1998PhRvC..58.1804A} and SLy \citep{2001A&A...380..151D} EoSs.}
\label{fig:b}
\end{figure}

\begin{figure}[!]
\centering
\includegraphics[scale=0.5]{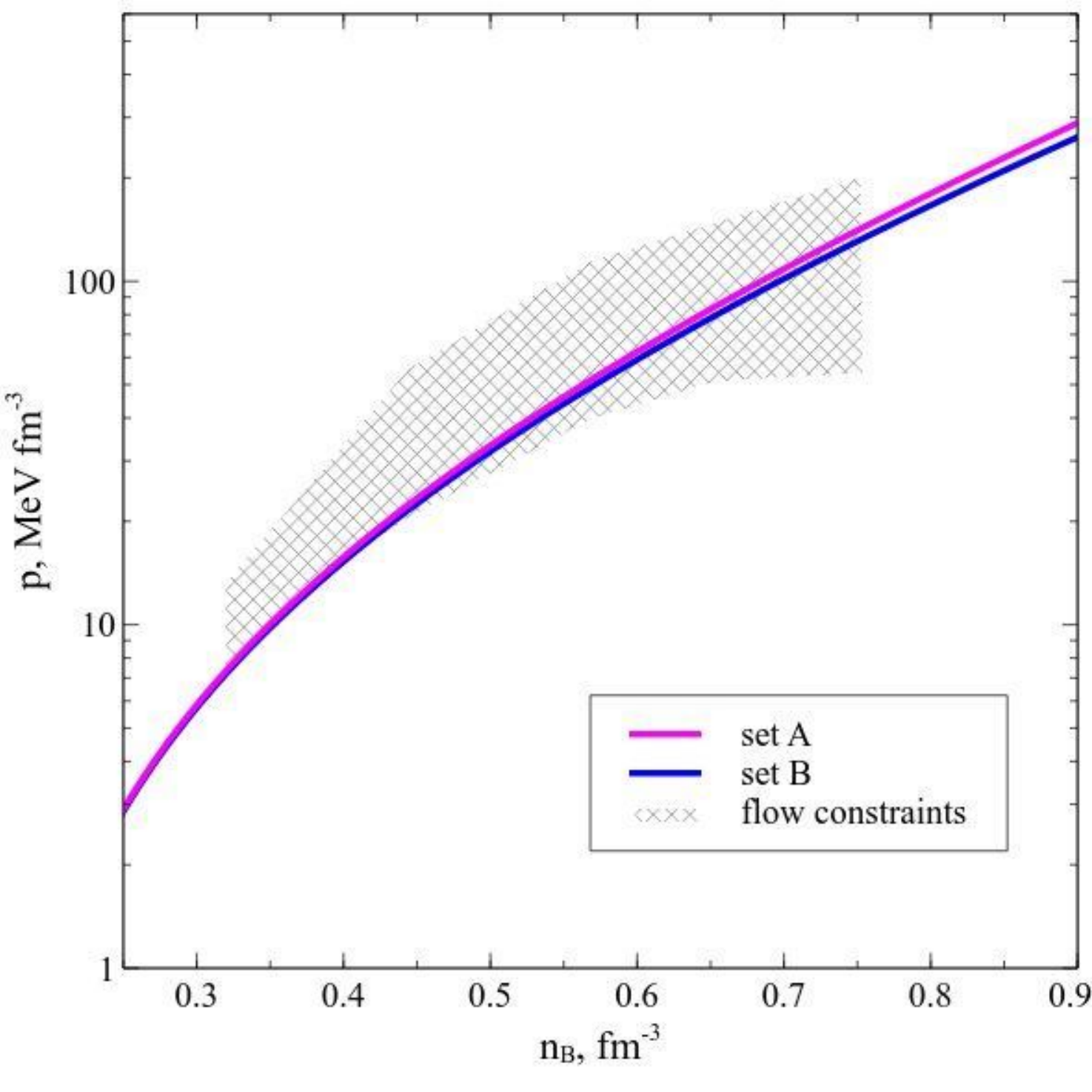}  
\caption{System pressure $p$ as a function of baryonic density $n_B$ for symmetric nuclear matter for the parameter sets from Table \ref{tab1}. The shaded area corresponds to the proton flow constraint \citep{2002Sci...298.1592D}.
}
\label{fig:c}
\end{figure}

\begin{figure}[!]
\centering
\includegraphics[scale=0.5]{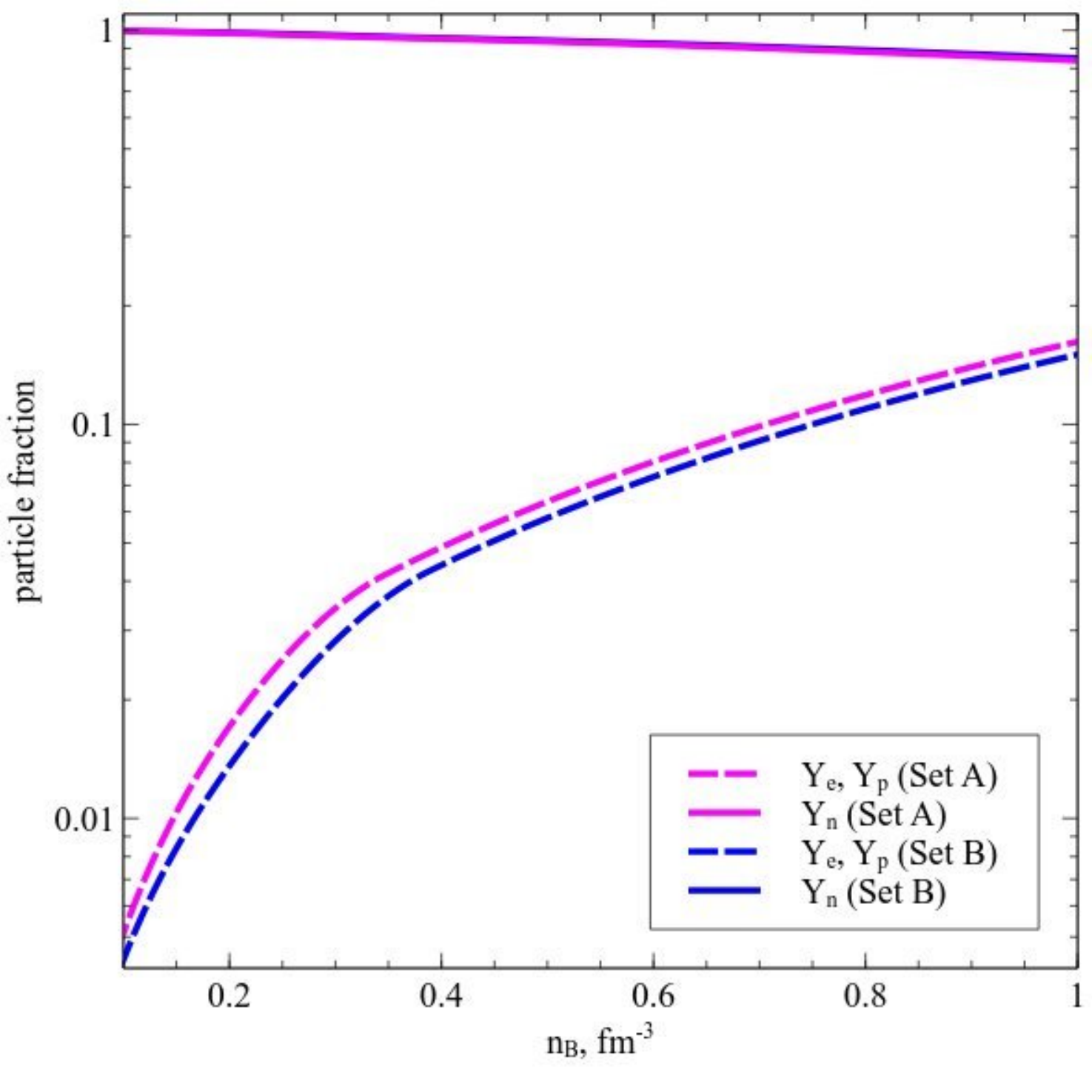}  
\caption{Fraction of electrons $Y_e$, protons $Y_p$ and neutrons $Y_p$ as functions of baryonic density $n_B$ for the parameter sets A and B from Table \ref{tab1}. The fractions of electrons and protons coincide due to electric neutrality.
}
\label{fig:d}
\end{figure}

The  mass-radius relations computed for the model parameter sets A and B are shown on the upper panel of Fig.~\ref{fig:b}. Both curves are in full agreement with the maximal observed NS mass equal to 2.01(4) $M_{\odot}$ \citep{2013Sci...340..448A}. The magenta curve (set A) lies in the region constrained by the astrophysical observational data ~\citep{2010ApJ...722...33S, 2013ApJ...765L...5S, 2016ARA&A..54..401O} and the statistically most probable configurations for any value of the stellar mass found in \citet{2018arXiv180300549M}. At the same time, the blue curve (set B), which is chosen to be consistent with the mass-radius relations of the APR4 and SLy EoSs, corresponds to the smaller star radii and lower maximal value of NS mass. As an additional quantity, we have also computed the NS compactness for the IST EoS with parameter sets A and B as well as for the APR4 and SLy EoSs. It is seen from the lower panel of Fig.\ref{fig:b} that these EoSs have nearly identical compactness for all values of the NS mass. The Set B represents almost the same compactness as the APR4 and SLy EoSs, even for $M$ close to the highest NS mass. This is interesting because such a regime corresponds to the highest baryonic density inside the NSs, for which the IST effects are the most important. We hope future NS mergers will provide the possibility of testing the validity of these EoSs. We also stress that, as seen from Fig. \ref{fig:c}, both parameterizations of the IST EoS are in full agreement with the proton flow constraint for all available densities ~\citep{2002Sci...298.1592D}. 

The previous facts motivate us to assume that both parameterizations of the IST EoS (with the A and B sets of parameters from Table \ref{tab1}) provide equally realistic descriptions of matter inside the NS. Because the IST EoS with parameter set B gives a similar mass-radius relation and nearly the same compactness as the APR4 and SLy EoSs, we expect the present EoS to be in good agreement with the NS-NS binary merger observations~\citep{2017PhRvL.119p1101A}. 

Finally, the IST EoS with the parameter set A (magenta curve in Fig. \ref{fig:b}) satisfies all the mentioned constraints and lies inside the most likelihood region of the parameter space (shaded cyan area in the upper panel of Fig. \ref{fig:b}). Therefore, we consider this set to be the best parameterization of the present EoS. We hope that further NS observations will help us to parameterize the IST EoS with even higher reliability.

This comparative study of two sets of IST EoS allows us to draw some generic conclusions: the nucleon hard core radius defines the stiffness of the EoS as first shown by~\citet{2017arXiv170907898S}, the parameter $\alpha$ defines the upper part of the mass-radius relation curve, the reduction of  $B^{sym}$  shifts the radius of the NSs to higher values, and the attraction term, which is accounted for via the mean field potential $U(n_{B}^{id})$, defines the part of the NS mass-radius relation that corresponds to the low masses and high radii. 

In this study, we found that the optimal value of $\alpha$ corresponds to $1.245$ (set A in Table \ref{tab1}). This value is consistent with previous results for the NS properties ~\citep{2017arXiv170907898S}, and hadron production in heavy-ion collisions~\citep{2017EPJWC.13709007S}. It is remarkable that this value corresponds to the Boltzmann statistics case. On the other hand, modeling of the NS interiors requires quantum statistics to be considered. This means that determination of $\alpha$ from the analysis of existing astrophysical constraints not only allows one to define this parameter for the quantum case but also constrains the possible values of the higher quantum virial coefficients, which remain unknown at the moment because they require many particle wave functions ~\citep{1996Pathria}.

Nevertheless, both parameterizations of the IST EoS (sets A and B) provide rather similar descriptions of the astrophysical data and give similar predictions of the NS chemical composition. This is illustrated by Fig. \ref{fig:d}, where fractions of neutrons, protons, and electrons are shown as a functions of baryonic density. It is clearly seen that sets A and B yield fractions of protons (and electrons) in the NS central area at $n_{B}=0.83 ~fm^{-3}$ of 12.5 \% and 11.5 \%, respectively. This result is in full accordance with the rapid cooling of the NS in MXB 1659-29.
As was shown by \citet{2018PhRvL.120r2701B} to describe the observational data of MXB 1659-29 the direct Urca reactions in the NS core have to occur. This requires the proton fraction to exceed a critical value $Y_{p,e}=0.11$, which is the case for both sets of IST EoS. 

\section{Summary and Conclusions}
\label{sec-4}
In this work we studied the NS properties within an elaborate approach that accounts for the asymmetry between neutrons and protons, as well as the effects of the IST coming from interparticle interaction. 
The star structure was assumed to include a core and an envelope corresponding to the outer and inner crusts of the NS.  The NS core was modeled within the IST EoS, while its crust was described via the polytropic EoS with $\gamma=4/3$. Our analysis shows that the crust-core transition occurs at the baryon densities much lower than the normal nuclear one and  its details affect only the tail of the mass-radius relation below the $0.2 M_{\odot}$. 
We found out a set of the IST EoS parameters which, on one hand, is consistent with astrophysical data, and, on other hand, reproduces the properties of normal nuclear matter, and agrees with the proton flow constraint. This EoS is also in full agreement with the experimental values of incompressibility factor, symmetry energy, and its slope at normal nuclear density. For both sets of parameters the IST EoS is in full accordance with the rapid cooling of the NS in MXB 1659-29. Furthermore, it is shown that in contrast to other hadronic EoSs with hard core repulsion, the contribution of the IST essentially softens  the EoS, and enables it to stay causal up to the baryonic density $0.83 ~fm^{-3}$ (5.2$n_{0}$), where the hadronic description must be replaced by the quark-gluon plasma one. 

Analysis of the NS mass-radius relation and the star compactness performed for the present EoS, and the APR4 and SLy EoSs, allows us to conclude that the IST EoS is able to describe the NS-NS binary merger data. This simple fact shows that the IST term plays a fundamental role in the EoS, and it is necessary for a correct description of the NS interiors. 

Finally, the presented IST EoS provides an elegant way to describe the properties of matter across a very wide range of densities and temperatures. In particular, the present model is applicable for descriptions of high- and low- energy nuclear collisions that establish unified treatment of nuclear matter at low densities and low temperatures, and of the hadron matter at low densities and high temperatures. Equally, the same EoS describes the strongly interacting matter inside the NSs at high density and vanishing temperatures.
Further development of the model will require accounting for heavy baryons and muons, studying the effects of the nuclear matter clusterization and testing the IST EoS on the core-collapse supernovae and NS merger observational data.

\begin{acknowledgments}
\vspace*{0.1cm}
The authors are grateful to the anonymous referee for the valuable comments and suggestions that improved significantly the contents and presentation of the article. V.S. and I.L. thank the Funda\c c\~ao para a Ci\^encia e Tecnologia (FCT), Portugal, for the financial support to the Centro de Astrof\'{\i}sica e Gravita\c c\~ao (CENTRA), Instituto Superior T\'ecnico, Universidade de Lisboa
through the grant No. UID/FIS/00099/2013. The work of A.I. was performed within the project SA083P17 of Universidad de Salamanca launched by the Regional Government of Castilla y Leon and the European Regional Development Fund. V.S. and A.I. also acknowledge partial support from The National Academy of Sciences of Ukraine (project No. 0118U003197).

\end{acknowledgments}
  
\bibliographystyle{yahapj}

\begin{thebibliography} 

\bibitem[Abbott et al. (2017)]{2017PhRvL.119p1101A} Abbott, B.~P. et al. (LIGO Scientific Collaboration and Virgo Collaboration)\ 2017,
\href{http://adsabs.harvard.edu/abs/2017PhRvL.119p1101A}{\prl, 119, 161101}

\bibitem[Akmal et al. (1998)]{1998PhRvC..58.1804A} Akmal, A., Pandharipande, V.~R., Ravenhall, D.~G.\ 1998, 
\href{http://adsabs.harvard.edu/abs/1998PhRvC..58.1804A}{\prc, 58, 1804}


\bibitem[Andronic et al. (2006)]{2006NuPhA.772..167A} Andronic, A., Braun-Munzinger, P., Stachel, J.\ 2006, 
\href{http://adsabs.harvard.edu/abs/2006NuPhA.772..167A}{\nphysa, 772, 167}

\bibitem[Andronic et al. (2017)]{2017JPhCS.779a2012A} Andronic, A., Braun-Munzinger, P., Redlich, K., et al.\ 2017, 
\href{http://adsabs.harvard.edu/abs/2017JPhCS.779a2012A}{\jphcs, 779, 012012}
	
\bibitem[Antoniadis et al. (2013)]{2013Sci...340..448A}  Antoniadis,  J., Freire,  P.~C.~C., N. Wex, et al.\ 2013, 
\href{http://adsabs.harvard.edu/abs/2013Sci...340..448A}{\sci, 340, 448}




\bibitem[Bao et al. (2014)]{2014PhRvC..90d5802B} Bao, S.~S., Hu, J.~N., Zhang, Z.~W., et al.\ 2014,
\href{http://adsabs.harvard.edu/abs/2014PhRvC..90d5802B}{\prc, 90, 045802}
	

\bibitem[Baym et al. (2018)]{2018RPPh...81e6902B} Baym, G., Hatsuda, T., Kojo, T., et al.\ 2018,
\href{http://adsabs.harvard.edu/abs/2018RPPh...81e6902B}{\rpph, 81, 056902}



	
\bibitem[Bethe (1971)]{1971ARNPS..21...93B} Bethe, H.~A.\ 1971,
\href{http://adsabs.harvard.edu/abs/1971ARNPS..21...93B}{\annurev, 21, 93}

\bibitem[Brown et al. (2018)]{2018PhRvL.120r2701B} Brown, E.~F., Cumming, A., Fattoyev, F.~J., et al. \ 2018,
\href{http://adsabs.harvard.edu/abs/2018PhRvL.120r2701B}{\prl, 120, 18270}

\bibitem[Bugaev  \& Gorenstein (1989)]{1989ZPhyC..43..261B} Bugaev, K.~A., Gorenstein, M.~I.\ 1989,
\href{http://adsabs.harvard.edu/abs/1989ZPhyC..43..261B}{\zphyc, 43, 261}
	
\bibitem[Bugaev	et al. (2018a)]{2017arXiv170406846B} Bugaev K.~A., Ivanytskyi  A.~I., Sagun V.~V., et al.\ 2018a, 
\href{http://adsabs.harvard.edu/abs/2017arXiv170406846B}{\ujp, 63, 863} 

\bibitem[Bugaev et al. (2018b)]{2018NuPhA.970..133B}
Bugaev, K.~A., Sagun, V.~V., Ivanytskyi, A.~I., et al.\ 2018b,
\href{http://http://adsabs.harvard.edu/abs/2018NuPhA.970..133B}{\nphysa, 970, 133}

\bibitem[Camelio et al. (2017)]{2017PhRvD..96d3015C} Camelio, G., Lovato, A., Gualtieri, L., et al.\ 2017, 
\href{http://adsabs.harvard.edu/abs/2017PhRvD..96d3015C}{\prd, 96, 043015}

\bibitem[Danielewicz et al. (2002)]{2002Sci...298.1592D} Danielewicz, P., Lacey, R., Lynch, W.~G.\ 2002, 
\href{http://adsabs.harvard.edu/abs/2002Sci...298.1592D}{\sci, 198, 1592}

%
\bibitem[Demorest et al. (2010)]{2010Natur.467.1081D} Demorest, P.~B., Pennucci, T., Ransom, S.~M.\ 2010,
\href{http://adsabs.harvard.edu/abs/2010Natur.467.1081D}{\nat, 467, 1081}

\bibitem[Douchin  \& Haensel (2001)]{2001A&A...380..151D} Douchin, F., \& Haensel, P.\ 2001,
\href{http://adsabs.harvard.edu/abs/2001A&A...380..151D}{\aap, 380, 151}
	
\bibitem[Dutra et al. (2012)]{2012PhRvC..85c5201D} Dutra, M., Louren{\c c}o, O., S{\'a} Martins, J.~S., et al.\ 2012,
\href{http://adsabs.harvard.edu/abs/2012PhRvC..85c5201D}{\prc, 85, 035201}


\bibitem[Hotokezaka et al. (2011)]{2011PhRvD...83..124008H} Hotokezaka, K., Kyutoku, K., Okawa, H., et al.\ 2011,
\href{http://adsabs.harvard.edu/abs/2011PhRvD..83l4008H}{\prd, 83, 124008} 

\bibitem[Ivanytskyi et al. (2018)]{2017arXiv171008218I} Ivanytskyi, A.~I., Bugaev, K.~A., Sagun, V.~V.,  et al.\ 2018, 
\href{http://adsabs.harvard.edu/abs/2017arXiv171008218I}{\prc, 97, 064905} 


\bibitem[Khan (2009)]{2009PhRvC..80a1307K} Khan, E. \ 2009, 
\href{http://adsabs.harvard.edu/abs/2009PhRvC..80a1307K}{\prc, 80, 011307}

\bibitem[Kh{\"a}n et al. (2006)]{2006PhRvC..74c5802K} Kl{\"a}hn, T., Blaschke, D., Typel, S., et al \ 2006, 
\href{http://adsabs.harvard.edu/abs/2006PhRvC..74c5802K}{\prc, 74, 035802}


\bibitem[Kolomeitsev et al. (2017)]{2017NuPhA.961..106K} 
Kolomeitsev, E. E., Maslov, K. A., Voskresensky, D. N. \ 2017, 
\href{http://adsabs.harvard.edu/abs/2017NuPhA.961..106K}
{\nphysa, 961, 106}


	

\bibitem[Lattimer (2012)]{2012ARNPS..62..485L} Lattimer, J.~M. \ 2012,
\href{http://adsabs.harvard.edu/abs/2012ARNPS..62..485L}{\arnps, 62, 485}

\bibitem[LIGO collab. (2017)]{2017ApJ...848L..12A} LIGO collaboration \ 2017,
\href{http://adsabs.harvard.edu/abs/2017ApJ...848L..12A}{\apjl, 848, L12}

\bibitem[Most et al. (2018)]{2018arXiv180300549M} Most, E.~R., Weih, L.~R., Rezzolla, L., et al.\ 2018,
\href{http://adsabs.harvard.edu/abs/2018arXiv180300549M}{\prl, 120, 261103}

\bibitem[Oliinychenko et al. (2012)]{2012arXiv1204.0103O} Oliinychenko, D.~R., Bugaev, K.~A., Sorin, A.~S.\ 2012,
\href{http://adsabs.harvard.edu/abs/2012arXiv1204.0103O}{\ujp, 58, 3, 211}

\bibitem[Oppenheimer  \& Volkoff (1939)]{1939PhRv...55..374O} Oppenheimer, J.~R., Volkoff, G.~M.\ 1939,
\href{http://adsabs.harvard.edu/abs/1939PhRv...55..374O}{\pr, 55, 374}

\bibitem[{\"O}zel  \&  Freire (2016)]{2016ARA&A..54..401O} {\"O}zel, F., \& Freire, P.\ 2016,
\href{http://adsabs.harvard.edu/abs/2016ARA&A..54..401O}{\aap, 54, 401}

\bibitem[Pathria (1996)]{1996Pathria} Pathria R. K. \ 1996,
\href{https://doi.org/10.1016/B978-0-7506-2469-5.X5000-2}{\stmec, 576}

\bibitem[Potekhin et al. (2013)]{2013A&A...560A..48P} Potekhin, A.~Y., Fantina, A.~F., Chamel, N., et al.\ 2013, 
\href{http://adsabs.harvard.edu/abs/2013A&A...560A..48P}{\aap, 560, 48}

	
\bibitem[Rischke et al. (1988)]{1988JPhG...14..191R} Rischke, D.~H., Friman, B.~L., St\"ocker, H., et al.\ 1988,
\href{http://adsabs.harvard.edu/abs/1988JPhG...14..191R}{\jphysg, 14, 191}

\bibitem[Rischke et al. (1991)]{1991ZPhyC..51..485} Rischke, D.~H., Gorenstein, M.~I., St\"ocker, H., et al.\ 1991,
\href{https://link.springer.com/article/10.1007/BF01548574}{\zphyc, 51, 485}
	

\bibitem[Sagun et al. (2017)]{2017EPJWC.13709007S} Sagun, V.~V., Bugaev, K.~A., Ivanytskyi, A.~I.,  et al.\ 2017, 
\href{http://adsabs.harvard.edu/abs/2017EPJWC.13709007S}{\epjwc, 137, 09007}

\bibitem[Sagun et al. (2018)]{2017arXiv170300049S} Sagun, V.~V., Bugaev, K.~A., Ivanytskyi, A.~I., et al.\ 2018, 
\href{http://adsabs.harvard.edu/abs/2017arXiv170300049S}{\epja, 54, 100}

	
\bibitem[Sagun et al. (2014)]{2014NuPhA.924...24S} Sagun, V.~V., Ivanytskyi, A.~I., Bugaev, K.~A., et al.\ 2014, 
\href{http://adsabs.harvard.edu/abs/2014NuPhA.924...24S}{\nphysa, 924, 24}


\bibitem[Sagun \& Lopes (2017)]{2017arXiv170907898S} Sagun, V.~V. \&  Lopes I.\ 2017, 
\href{https://doi.org/10.3847/1538-4357/aa92cf}{\apj, 850, 75}


\bibitem[Satz (2012)]{2012LNP...841.....S} Satz, H.\ 2012,
\href{http://adsabs.harvard.edu/abs/2012LNP...841.....S}{\lnp, 841, 29}

	

\bibitem[Shibata  \& Taniguchi (2011)]{2011LRR....14....6S} Shibata, M., Taniguchi, K.\ 2011.\
\href{http://adsabs.harvard.edu/abs/2011LRR....14....6S}{\ Living Reviews in Relativity 14, 6.}
	
\bibitem[Steiner et al. (2010)]{2010ApJ...722...33S} Steiner, A.~W., Lattimer, J.~M., Brown, E.~F.\ 2010
\href{http://adsabs.harvard.edu/abs/2010ApJ...722...33S}{\apj, 722, 33}
	
\bibitem[Steiner et al. (2013)]{2013ApJ...765L...5S} Steiner, A.~W., Lattimer, J.~M., Brown, E.~F.\ 2013
\href{http://adsabs.harvard.edu/abs/2013ApJ...765L...5S}{\apj, 765, 5}

%
\bibitem[Tolman (1934)]{1934rtc..book.....T} Tolman, R.~C.\ 1934, 
\href{http://adsabs.harvard.edu/abs/1934rtc..book.....T}{\tc, 234} 

%
\bibitem[Tolman (1939)]{1939PhRv...55..364T} Tolman, R.~C.\ 1939, 
\href{http://adsabs.harvard.edu/abs/1939PhRv...55..364T}{\pr, 55, 364}


\bibitem[Zhang  \& Chen (2013)]{2013PhLB..726..234Z} Zhang, Z., Chen, L.-W., \ 2013, 
\href{http://adsabs.harvard.edu/abs/2013PhLB..726..234Z}{\plb, 726, 234}

\end{thebibliography}

\end{document}